\documentclass[preprint,11pt]{elsarticle}

\usepackage[utf8]{inputenc}
\usepackage[T1]{fontenc}
\usepackage{amsmath,amssymb}
\usepackage{graphicx}
\usepackage{booktabs}
\usepackage{siunitx}
\usepackage{xcolor}
\usepackage[colorlinks=true,linkcolor=blue!50!black,citecolor=blue!50!black,urlcolor=blue!50!black]{hyperref}
\graphicspath{{figures/}}

\journal{Journal of Sound and Vibration}
\makeatletter\def\ps@pprintTitle{\let\@oddhead\@empty\let\@evenhead\@empty\let\@oddfoot\@empty\let\@evenfoot\@empty}\makeatother

\begin{document}

\begin{frontmatter}

\title{Virtual-array operational modal analysis of a rolling tire using a single tire cavity accelerometer}

\author[vt]{Pradosh P. Dash}
\author[tamu]{Pablo A. Tarazaga}
\affiliation[vt]{organization={Department of Mechanical Engineering, Virginia Polytechnic Institute and State University},
            city={Blacksburg}, state={VA}, postcode={24061}, country={USA}}
\affiliation[tamu]{organization={J. Mike Walker '66 Department of Mechanical Engineering, Texas A\&M University},
            city={College Station}, state={TX}, postcode={77843}, country={USA}}

\begin{abstract}
The dynamics of the rolling tire govern structure-borne road noise below \SI{500}{\hertz}. Laser Doppler vibrometry (LDV) is the established laboratory reference for measuring rolling-tire modes. This paper presents a complementary approach that uses a single wireless tire cavity accelerometer (TCA) on the inner liner and a once-per-revolution encoder on the drum of a tire-on-drum rig fitted with a cleat. Because the drum-to-tire diameter ratio (2.74) is not an integer, successive cleat impacts find the TCA at different circumferential positions. The TCA's own contact-patch signature locates the sensor at each impact within about \SI{1}{\degree}, and grouping the impact responses by position synthesizes a 36-element virtual array from one \SI{128}{\second} run. Covariance-driven stochastic subspace identification is applied with a consensus rule over several settings, a bootstrap over impacts and a correction for window-induced damping, and the complete chain is verified on synthetic data with known modes. For a slick 195/65R15 tire at \SI{30}{\kilo\meter\per\hour} and \SI{1334}{\newton}, ten modes between 46 and \SI{193}{\hertz} were identified robustly, with circumferential wavenumbers 1 to 5, damping ratios of 2.1--4.2\,\% and strongly complex shapes; three further modes are tentative. The modes follow the wavenumber ladder of a stationary modal test of the same tire. The method needs no optical access and records the whole array in one run, which suits treaded tires and on-road testing, at the cost of installing a sensor inside the tire.
\end{abstract}


\end{frontmatter}

%% ===================================================================
\section{Introduction}
\label{sec:intro}

As engine noise falls in hybrid and electric vehicles, tire--pavement interaction has become a dominant source of interior noise \cite{govindswamy2009,chang2010}. Below about \SI{500}{\hertz} the tire contribution reaches the cabin mainly as structure-borne noise (SBN) through the wheel and suspension, while air-borne noise dominates at higher frequencies \cite{lopez2007}. Impacts of the tread and of road texture at the contact patch, and the deformation of the carcass as it enters and leaves the patch, excite flexural waves that travel around the belt \cite{sandberg2002}. Below \SI{500}{\hertz} these waves are weakly damped and interfere to form circumferential modes \cite{pinnington2002,pinnington2006}, which control the forces transmitted to the spindle. Predicting and mitigating SBN therefore requires the modal properties of the tire under realistic operating conditions.

Experimental modal analysis (EMA) of stationary tires, with shaker excitation and accelerometer or laser responses, is mature \cite{bolton1998,yam2000,kindt2006,rocca2011}. A spindle load splits the repeated circumferential modes of the axisymmetric tire into pairs, one with a node and one with an antinode at the contact patch \cite{kindt2006,wheeler2005}. Rolling changes the dynamics further: cyclic deformation softens the rubber, rotation introduces gyroscopic coupling, and the contact patch moves relative to the rotating structure \cite{kindt2009rolling,rocca2011indoor}. These effects can only be captured by measurements on the rolling tire, where the input force is not measured and output-only, or operational, modal analysis (OMA) is required \cite{james1995,brincker2015}.

Laser Doppler vibrometry is the established reference for such measurements. Kindt et al.\ \cite{kindt2009measurement,kindt2011} and Rocca et al.\ \cite{rocca2011indoor} measured the tread and sidewall of smooth tires rolling against a cleated drum or wheel, using a stationary LDV and mirrors placed sequentially around the tire, and identified rolling modes up to several hundred hertz. LDV is non-contact, leaves the tire unmodified and offers a fine, freely chosen measurement grid. Its main practical demands are an optical path to the tread, which for a loaded tire is usually provided by mirrors near the contact region, and a sequential scan whose duration grows with the number of points. Speckle and signal dropout on rough or dark rotating surfaces have historically limited LDV on treaded tires \cite{bell2000}; continued developments in laser sources, including infrared lasers that return more light from dark rubber, and in tracking and processing have reduced these difficulties \cite{rothberg2017}.

Sensors mounted inside the tire offer a different set of trade-offs. Burroughs et al.\ \cite{burroughs2003} bonded accelerometers to the tread liner to measure tread vibration, and in-tire accelerometers are now widely studied for estimating contact-patch and road-friction quantities in so-called intelligent tires \cite{lee2017intelligent}. Ishihama et al.\ \cite{ishihama2017} used a commercial wireless tire cavity accelerometer (TCA) to study tread vibration under driving force and road roughness. An in-tire sensor needs no optical access, works with any tread pattern and travels with the vehicle, but it must be installed inside the tire and it measures at a single point that rotates with the tire. To our knowledge, a single in-tire sensor has not been used to identify the circumferential modes of a rolling tire.

This paper shows that a single TCA can provide a full circumferential array on a standard tire-on-drum rig, and presents it as a practical complement to LDV rather than a replacement. The key observation is that the ratio between the drum and tire diameters is generally not an integer, so that the TCA occupies a different angular position each time the cleat strikes the tire. Time-stamping each impact and the passage of the TCA through the contact patch locates the sensor at every impact, and grouping the impact responses by position synthesizes a virtual array from one continuous run. The specific contributions are:
\begin{itemize}
\item the virtual-array measurement principle, which needs only the TCA and a once-per-revolution encoder on the drum;
\item a self-referenced angular origin taken from the contact-patch signature of the TCA itself, validated against an optical tire encoder;
\item a reproducible identification protocol for the short, highly damped impulse responses involved: a consensus over several covariance-driven stochastic subspace identification (SSI-Cov) settings, a bootstrap over impacts for uncertainty and detection rates, and a correction for the damping introduced by exponential windowing;
\item a verification of the complete chain on synthetic data with known modes and realistic noise, together with sensitivity studies of the processing choices;
\item a physical check of the identified modes against a stationary EMA of the same tire.
\end{itemize}
The experiments were part of the first author's MS thesis \cite{dash2020thesis}; all results presented here were obtained from the raw records with the processing described below, and supporting material is given in the Supplementary Information (SI).

The paper is organized as follows. Section~\ref{sec:principle} describes the measurement principle, the rig and the angular reference. Section~\ref{sec:processing} presents the signal conditioning and Section~\ref{sec:oma} the identification protocol and its verification. Results are given in Section~\ref{sec:results} and discussed, together with the practical trade-offs relative to LDV, in Section~\ref{sec:discussion}.

%% ===================================================================
\section{Measurement principle and experimental set-up}
\label{sec:principle}

\subsection{Virtual array from a non-integer diameter ratio}

\begin{figure}[t]
\centering
\includegraphics[width=\textwidth]{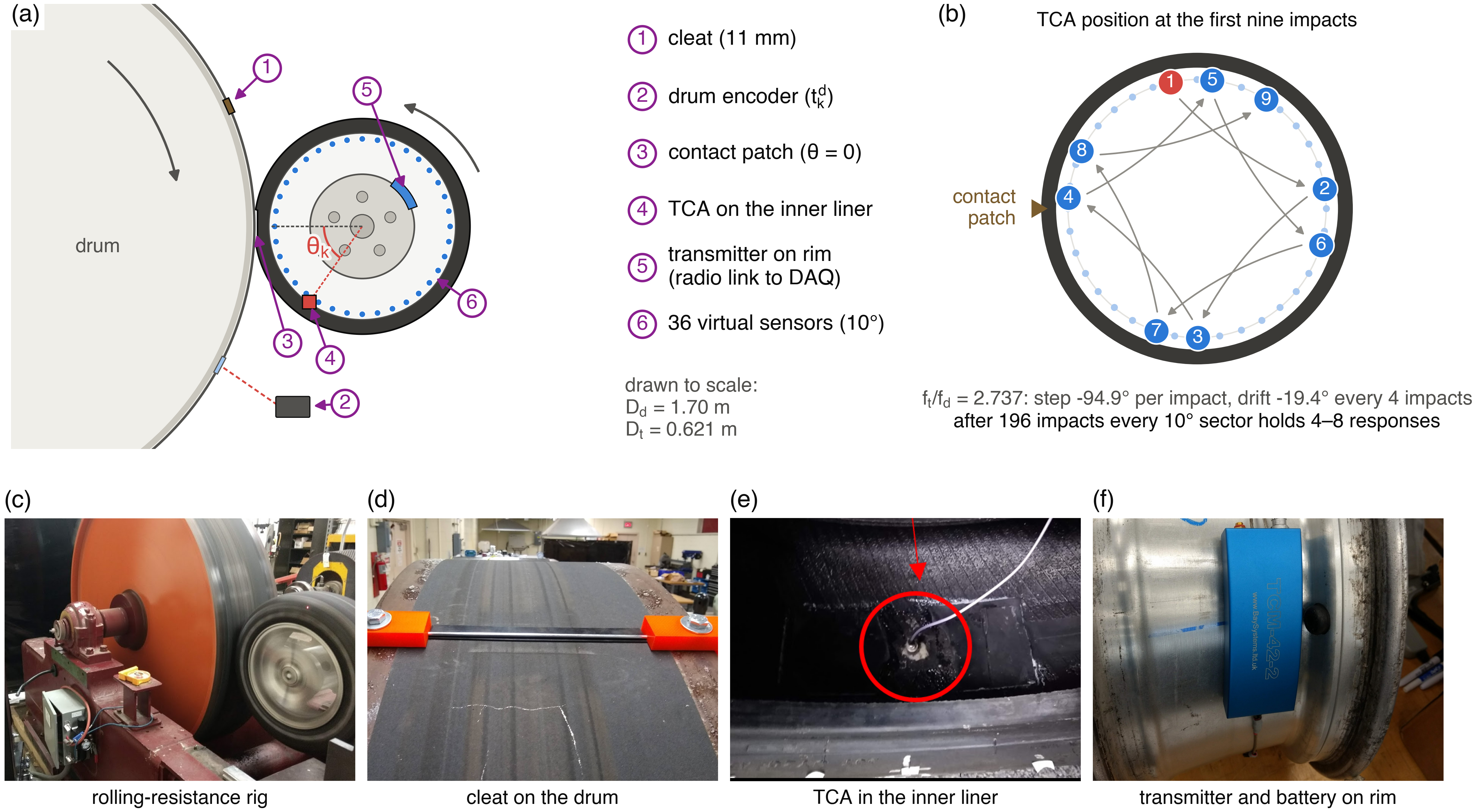}
\caption{Measurement concept and hardware. (a) Tire-on-drum configuration, drawn to scale; the numbered components are listed in the key. (b) Measured position of the TCA at the first nine cleat impacts of the test: each drum revolution moves the sensor by \SI{-94.9}{\degree} relative to the contact patch, and the slow drift of the pattern fills all 36 sectors. (c) Rolling-resistance rig at the Center for Tire Research (CenTiRe). (d) Cleat assembly on the drum. (e) TCA threaded into an insert in the inner liner at the tread center. (f) TCA transmitter and battery strapped to the rim.}
\label{fig:setup}
\end{figure}

Figure~\ref{fig:setup}(a) sketches the configuration. A cleat fixed to the drum strikes the tire once per drum revolution, always at the contact patch. Without slip at the tire--drum interface, the surface speeds match and the revolution frequencies of the tire, $f_\mathrm{t}$, and of the drum, $f_\mathrm{d}$, satisfy
\begin{equation}
r \equiv \frac{f_\mathrm{t}}{f_\mathrm{d}} = \frac{D_\mathrm{d}}{D_\mathrm{t}},
\label{eq:ratio}
\end{equation}
where $D_\mathrm{d}$ is the drum diameter and $D_\mathrm{t}$ the effective rolling diameter of the loaded tire. Between two impacts the tire turns $r$ times, so the TCA advances by
\begin{equation}
\Delta\theta = 360^\circ\,(r - \lfloor r \rfloor)
\label{eq:advance}
\end{equation}
relative to the contact patch. When $r$ is not an integer, and in particular when $r-\lfloor r\rfloor$ is not a simple fraction, the impact positions $\theta_k$ of the TCA drift around the circumference and, over enough drum revolutions, cover it densely (Fig.~\ref{fig:setup}b). Each impact response is then the response of the tire to an (ideally identical) cleat input, recorded at a different position. Grouping the responses by position and averaging within each group yields the responses of a set of virtual sensors distributed around the tire, as if a circumferential array had been installed.

The angular position at impact $k$ follows from two time stamps: the time $t^\mathrm{d}_k$ at which the cleat passes the contact patch, and the time $t^\mathrm{t}_{j}$ of the last passage of the TCA through the contact patch before it,
\begin{equation}
\theta_k = \mathcal{W}\!\left[\,360^\circ\, \bar f_\mathrm{t}\,\big(t^\mathrm{d}_k - t^\mathrm{t}_{j}\big) + \theta_\mathrm{ref}\right],
\label{eq:theta}
\end{equation}
where $\bar f_\mathrm{t}$ is the tire revolution frequency averaged over the five revolutions centered on the impact, $\theta_\mathrm{ref}$ is the angle between the tire time stamp and the center of the contact patch (Section~\ref{sec:reference}), and $\mathcal{W}$ wraps the angle to $(-180^\circ, 180^\circ]$. The angle is positive along the direction of rotation, so that $\theta=0$ is the center of the contact patch and $\theta=\pm180^\circ$ the point opposite to it.

\subsection{Test rig, tire and instrumentation}

The tests were run on a rolling-resistance rig at the Center for Tire Research (CenTiRe), Virginia Tech (Fig.~\ref{fig:setup}c). The drum has a diameter of \SI{1.70}{\meter} and a P120 abrasive surface. A slick radial tire of size 195/65R15 (unloaded diameter \SI{0.630}{\meter}, circumference \SI{1.984}{\meter}) was used, so that tread-impact excitation was absent. The tire was inflated to \SI{220.6}{\kilo\pascal} (32~psi) unloaded and run at zero slip angle. The cleat (Fig.~\ref{fig:setup}d) is a steel rod of \SI{6.5}{\milli\meter} diameter on a \SI{4.5}{\milli\meter} 60A neoprene pad, \SI{11}{\milli\meter} high in total, press-fitted into 3D-printed brackets bolted to the drum.

The TCA system comprises a uniaxial IEPE accelerometer (nominal sensitivity \SI{1.06}{\milli\volt\per g} at \SI{200}{\hertz}, $\pm5$\,\% from \SI{1}{\hertz} to \SI{10}{\kilo\hertz}), a transmitter and a battery, and a receiver connected to the data acquisition system. The accelerometer is threaded into a metal insert that is press-fitted and bonded into a hole in the inner liner at the tread center, and measures the radial acceleration (Fig.~\ref{fig:setup}e). The transmitter and battery are strapped to the rim and rotate with the wheel (Fig.~\ref{fig:setup}f). The insert locally modifies the liner; its effect on the global modes is assumed to be small \cite{dash2020thesis}. A retro-reflective photoelectric sensor (Banner SM312LVMHS) facing a reflective tape on the drum provides the cleat time stamps $t^\mathrm{d}_k$.

The results reported here are from one run at \SI{30}{\kilo\meter\per\hour} and a spindle load of \SI{1334}{\newton} (300~lbf), recorded for \SI{128}{\second} at \SI{2048}{\hertz} during the first of two cleat-test campaigns (dataset D1; tire temperature not recorded). The TCA signal was converted to acceleration with a sensitivity of \SI{1.02}{\milli\volt\per g}; the absolute scale does not affect the modal parameters. A second campaign (dataset D2: same tire, speed and load, a smaller cleat, \SI{300}{\second} at \SI{4096}{\hertz}) included an optical encoder on the tire sidewall and is used here only to validate the angular reference; its smaller cleat excited the tire too weakly for modal identification. The D1 record contains 199 drum and 542 tire revolutions. The median revolution frequencies are $f_\mathrm{d}=\SI{1.553}{\hertz}$ and $f_\mathrm{t}=\SI{4.249}{\hertz}$, so $r=2.737$ (Eq.~\ref{eq:ratio}), which corresponds to a rolling diameter $D_\mathrm{t}=\SI{0.621}{\meter}$, \SI{1.4}{\percent} below the unloaded diameter, and to an advance of \SI{265.1}{\degree} (equivalently \SI{-94.9}{\degree}) per drum revolution (Eq.~\ref{eq:advance}). Four impacts bring the sensor back to within \SI{19.4}{\degree} of its starting position, and this slow drift distributes the impacts over all sectors (Figs.~\ref{fig:setup}b and \ref{fig:va}b).

\subsection{Angular reference from the contact-patch signature}
\label{sec:reference}

\begin{figure}[t]
\centering
\includegraphics[width=\textwidth]{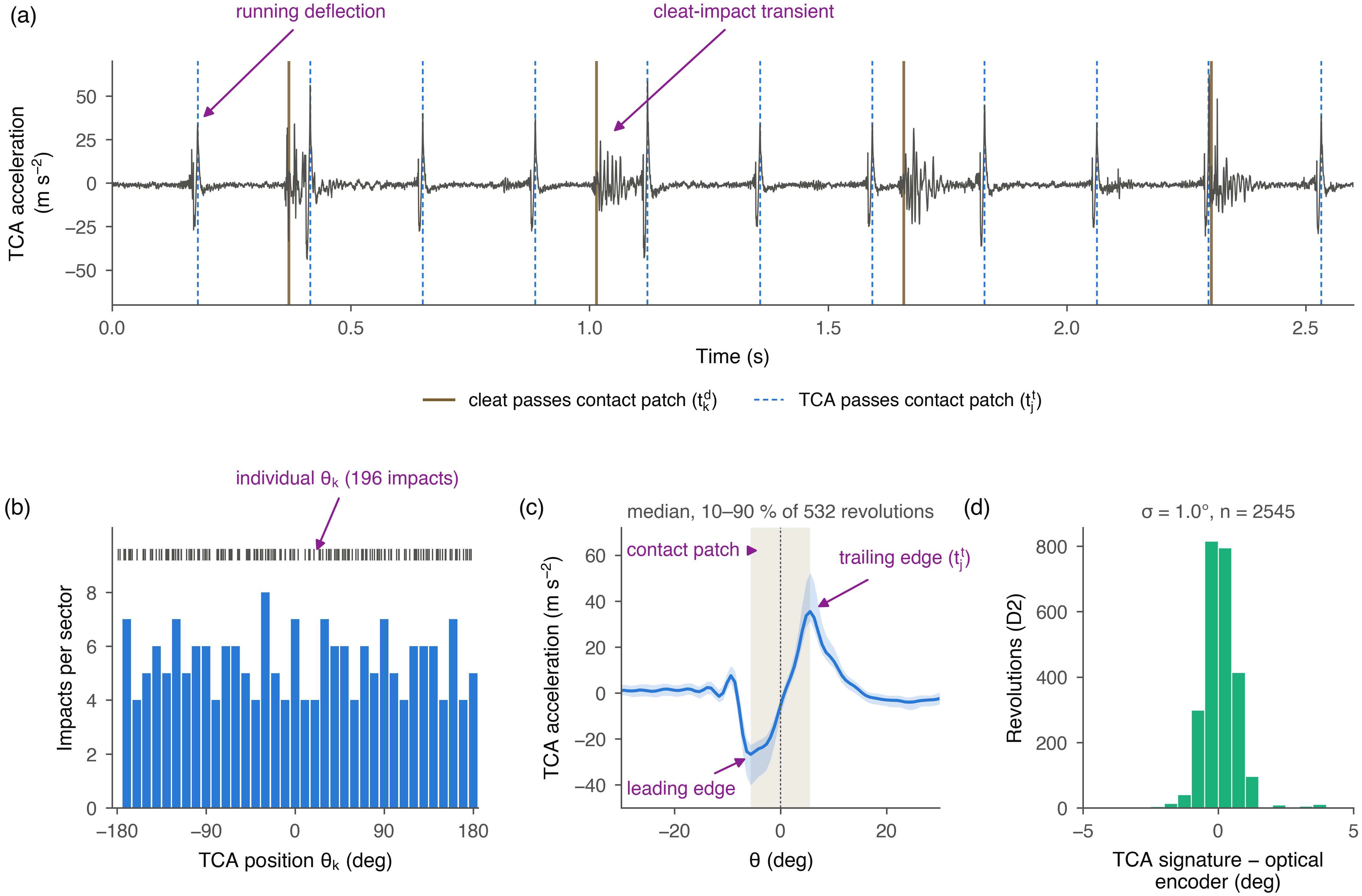}
\caption{Synthesis of the virtual array and its angular reference. (a) TCA signal of D1 over \SI{2.6}{\second} with the cleat time stamps from the drum encoder and the contact-patch passages of the TCA. (b) Number of impacts in each of the 36 sectors; ticks above the bars mark the individual impact positions $\theta_k$ ($\theta=0$: contact-patch center, positive along the rotation). (c) Contact-patch signature of the TCA (low-pass \SI{400}{\hertz}) that defines the angular origin. (d) Deviation of the signature midpoint from an optical tire encoder over 2545 revolutions of dataset D2. Annotations are drawn in purple.}
\label{fig:va}
\end{figure}

Each passage of the TCA through the contact patch produces a characteristic waveform with peaks of opposite sign at the leading and trailing edges \cite{sandberg2002}; in D1 it consists of a negative trough as the liner enters the patch and a positive peak as it leaves (Figs.~\ref{fig:va}a and \ref{fig:va}c; the polarity depends on the signal chain). In D1 the tire time stamps $t^\mathrm{t}_j$ were taken directly from this signature, as the positive (trailing-edge) peak of the TCA signal band-passed between 1 and \SI{150}{\hertz}; a peak was found in every one of the 542 revolutions. The synchronous average over 532 revolutions (Fig.~\ref{fig:va}c) places the leading-edge trough \SI{11.2}{\degree} before the trailing-edge peak. Taking the contact-patch center midway between the two gives $\theta_\mathrm{ref}=\SI{5.6}{\degree}$ in Eq.~(\ref{eq:theta}). The corresponding transit time of a liner point through the patch is $\tau_\mathrm{cp}=\SI{7.3}{\milli\second}$.

The precision of this self-referenced origin was checked on D2, where an optical encoder on the sidewall gave an independent tire time stamp. Over 2545 revolutions, the midpoint of the TCA signature relative to the optical mark had a standard deviation of \SI{1.0}{\degree} (Fig.~\ref{fig:va}d), which includes the jitter of the optical encoder itself and is an order of magnitude smaller than the \SI{10}{\degree} angular bins used below. The same comparison revealed a constant offset of \SI{42.6}{\degree} between the optical mark and the TCA (SI, Fig.~S5), most likely caused by the placement of the reflective tape relative to the sensor. Referencing the angle to the TCA signature avoids this kind of installation error and removes the need for a tire encoder: the method requires only the TCA and the drum encoder.

\subsection{Overview of the processing chain}
\label{sec:chain}

The processing that turns the single TCA record into modal parameters has eight steps, illustrated with the data of D1 in SI Fig.~S1. (i) The TCA signal and the drum encoder are recorded simultaneously. (ii) Each cleat passage $t^\mathrm{d}_k$ is time-stamped from the drum encoder and each passage of the TCA through the contact patch $t^\mathrm{t}_j$ from the TCA signature (Section~\ref{sec:reference}). (iii) The tire-periodic response to the running deflection is removed by order tracking, and (iv) the remaining signal is cut into one drum revolution per impact (Section~\ref{sec:processing}). (v) The position $\theta_k$ of the sensor at every impact follows from the two time stamps (Eq.~\ref{eq:theta}), and (vi) the impacts are grouped into 36 angular sectors, averaged and windowed to form the virtual-array matrix $\mathbf{Y}$. (vii) Covariance-driven stochastic subspace identification is applied to $\mathbf{Y}$ with several settings, and (viii) the modes are extracted by a consensus rule, their uncertainty is estimated by a bootstrap over impacts, and their damping is corrected for the window (Section~\ref{sec:oma}).

%% ===================================================================
\section{Signal conditioning}
\label{sec:processing}

\subsection{Removal of the running-deflection response}

The TCA signal contains two contributions of different nature (Fig.~\ref{fig:va}a): the transient response to the cleat, which carries the modal information, and the response to the running deflection of the tire, which repeats once per tire revolution. The latter was removed by order tracking with respect to the tire revolution \cite{feng2017,li2018}. Each tire revolution, delimited by consecutive time stamps $t^\mathrm{t}_j$, was resampled with an anti-aliasing polyphase filter to 450 points, that is, to a uniform angular grid. The resampled revolutions were averaged, which is equivalent to averaging their discrete Fourier coefficients and retains only the components phase-locked to the tire rotation. The average was resampled back to the length of each revolution and subtracted from the measured signal, leaving the non-periodic part (Fig.~\ref{fig:conditioning}a).

\subsection{Segmentation, clustering and averaging}

The non-periodic signal was cut into segments of one drum revolution, each starting at a cleat time stamp $t^\mathrm{d}_k$, and each segment was resampled to $N=1300$ points per drum revolution. This removes the small variations of drum speed and gives all segments a common sampling rate of $f_s = N f_\mathrm{d} = \SI{2018.5}{\hertz}$ and a duration of \SI{0.644}{\second}. Discarding the first impact (start-up) and the impacts without a complete tire revolution on either side left 196 impacts. Each impact was assigned to the nearest of 36 sector centroids spaced \SI{10}{\degree} apart, the first one at the contact-patch center. Every sector received between 4 and 8 impacts (median 5; Fig.~\ref{fig:va}b).

Averaging within a sector assumes that the cleat input is repeatable. The data support this assumption (SI, Fig.~S4): segments of the same sector have a median correlation of 0.66 in the first \SI{0.1}{\second} (30--\SI{220}{\hertz}), against $-0.08$ for segments of sectors at least \SI{90}{\degree} apart, and the impact energy in the first \SI{30}{\milli\second} varies by only 7\,\% (median coefficient of variation) within a sector, while it varies by 26\,\% around the tire.

Because the segments are synchronized with the drum, the residual contact-patch passages that survive the order tracking are not aligned from one segment to the next (Fig.~\ref{fig:conditioning}c), so the sector average was multiplied by an exponential window
\begin{equation}
w_n = \exp\!\left(-\frac{n}{N-1}\ln\frac{100}{\beta}\right), \qquad n = 0,\dots,N-1,
\label{eq:window}
\end{equation}
with end factor $\beta = 1$\,\% to attenuate these residuals and limit leakage. The result is the virtual-array matrix $\mathbf{Y}\in\mathbb{R}^{N\times 36}$, whose columns are the averaged, windowed impulse responses at the 36 virtual sensors.

The exponential window adds a known decay rate
\begin{equation}
\sigma_w = \frac{f_s}{N-1}\ln\frac{100}{\beta} = \SI{7.16}{\per\second}
\label{eq:sigmaw}
\end{equation}
to every mode in $\mathbf{Y}$. If $\lambda = -\zeta_\mathrm{id}\omega_n + \mathrm{i}\,\omega_d$ is an identified pole, the damping of the unwindowed response is recovered as \cite{fladung1997}
\begin{equation}
\zeta = \frac{\zeta_\mathrm{id}\,\omega_n - \sigma_w}{\sqrt{(\zeta_\mathrm{id}\,\omega_n - \sigma_w)^2 + \omega_d^2}} \;\approx\; \zeta_\mathrm{id} - \frac{\sigma_w}{\omega_n}.
\label{eq:zcorr}
\end{equation}
At \SI{50}{\hertz} the correction amounts to 2.3 percentage points of damping, at \SI{190}{\hertz} to 0.6.

\subsection{Character of the conditioned signals}

\begin{figure}[t]
\centering
\includegraphics[width=\textwidth]{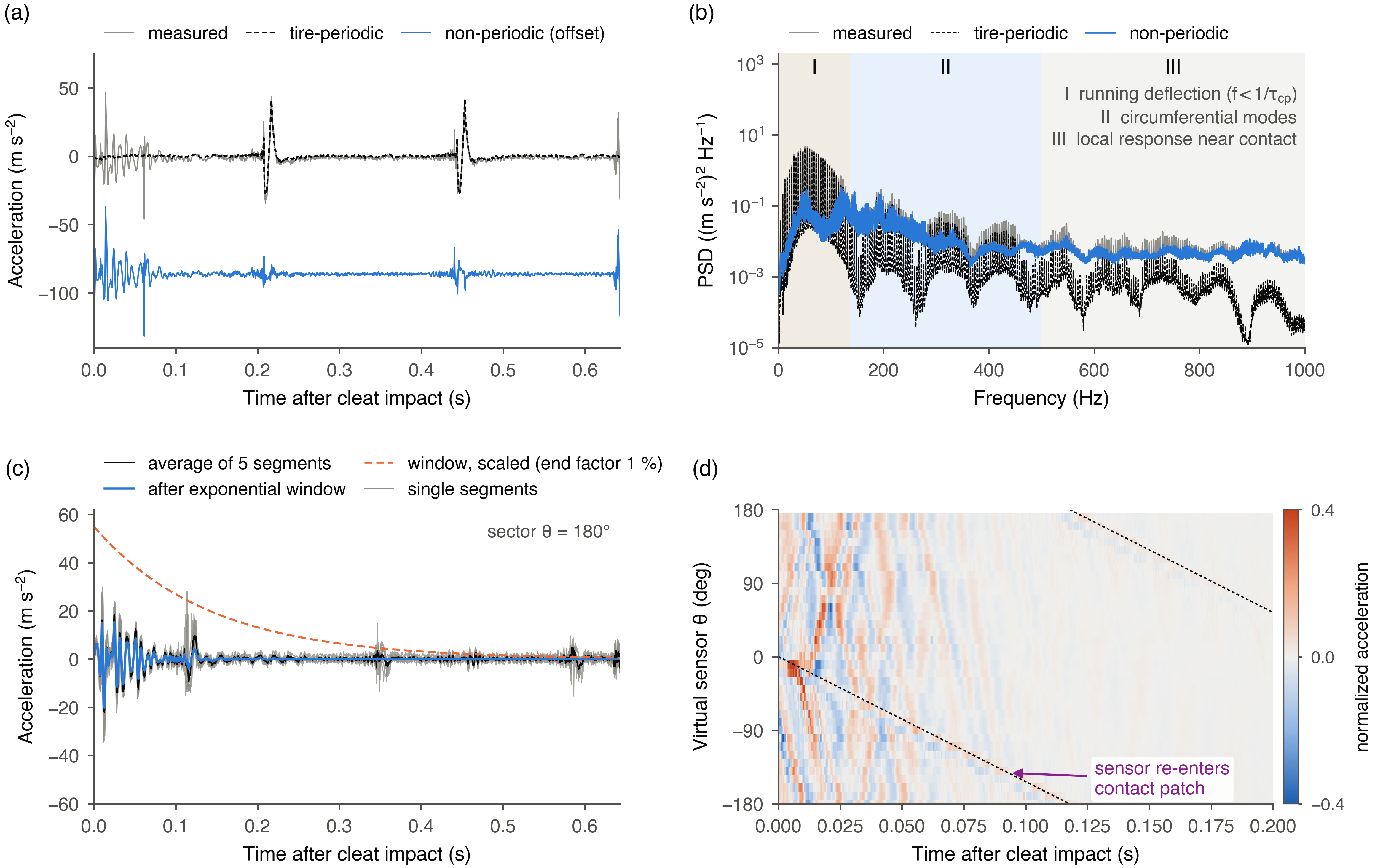}
\caption{Signal conditioning (dataset D1). (a) One drum revolution after a cleat impact: measured TCA signal, tire-periodic component from order tracking, and the non-periodic remainder (offset for clarity). (b) Power spectral densities of the whole record, with the three frequency regions; $\tau_\mathrm{cp}$ is the contact-patch transit time from Fig.~\ref{fig:va}(c). (c) The five segments of the sector at $\theta=180^\circ$, their average, the exponential window of Eq.~(\ref{eq:window}) (scaled) and the windowed average. (d) Space--time map of the virtual-array matrix $\mathbf{Y}$ (normalized); the dashed line marks the predicted re-entry of each virtual sensor into the contact patch.}
\label{fig:conditioning}
\end{figure}

The spectra in Fig.~\ref{fig:conditioning}(b) show three regions. Below about \SI{140}{\hertz} the measured signal is dominated by the tire-periodic component, a dense comb of tire orders whose envelope is set by the duration of the contact-patch event; the upper limit of this region agrees with $1/\tau_\mathrm{cp} = \SI{137}{\hertz}$ obtained from Fig.~\ref{fig:va}(c), consistent with a forced response to the static part of the running deflection \cite{molisani2003}. Between about 140 and \SI{500}{\hertz} the non-periodic part carries most of the energy and the response persists around the tire (Fig.~\ref{fig:conditioning}d), which is characteristic of the circumferential modes \cite{pinnington2002,pinnington2006}. Above \SI{500}{\hertz} the vibration is confined to the vicinity of the contact patch and decays before it can travel around the tire. Between 20 and \SI{130}{\hertz} the order tracking removes the tire-order harmonics, which stand up to about \SI{13}{\decibel} above the non-periodic level, and lowers the total power in this band by a factor of five (\SI{7}{\decibel}); this is what makes the low-frequency modes accessible.

Figure~\ref{fig:conditioning}(d) displays $\mathbf{Y}$ as a space--time map. The response is strongest during the first \SI{0.05}{\second} after the impact and decays within about \SI{0.1}{\second}. The narrow oblique trace is the residual contact-patch passage of each virtual sensor: a sensor at $\theta<0$ reaches the patch after $|\theta|/(360^\circ f_\mathrm{t})$, which is the dashed line.

%% ===================================================================
\section{Modal identification}
\label{sec:oma}

\subsection{Frequency domain decomposition}

As a non-parametric reference, the cross-power spectral density matrix of the 36 columns of $\mathbf{Y}$ was estimated with Welch's method (Hann window of 512 samples, 75\,\% overlap, resolution \SI{3.9}{\hertz}) and decomposed by singular value decomposition at each frequency \cite{brincker2001}. Peaks of the first singular values indicate modes; in the frequency domain decomposition (FDD) the corresponding singular vectors estimate the mode shapes.

\subsection{Covariance-driven stochastic subspace identification}

SSI-Cov \cite{vanOverschee1996,peeters1999} fits a discrete-time state-space model to the output covariances. With $\mathbf{y}_t\in\mathbb{R}^{36}$ the $t$-th row of $\mathbf{Y}$, the covariances of the mean-removed responses, $\mathbf{R}_i = \frac{1}{N}\sum_t \mathbf{y}_{t+i}\mathbf{y}_t^{\mathsf T}$, for $i=1,\dots,2s$ are arranged in the block Hankel matrix
\begin{equation}
\mathbf{H} = \begin{bmatrix} \mathbf{R}_1 & \mathbf{R}_2 & \cdots & \mathbf{R}_s \\ \mathbf{R}_2 & \mathbf{R}_3 & \cdots & \mathbf{R}_{s+1} \\ \vdots & & & \vdots \\ \mathbf{R}_s & \mathbf{R}_{s+1} & \cdots & \mathbf{R}_{2s-1} \end{bmatrix} = \mathbf{U}\boldsymbol{\Sigma}\mathbf{V}^{\mathsf T},
\label{eq:hankel}
\end{equation}
where $s$ is the number of block rows. For a model order $n$, the observability matrix is $\mathbf{O}_n = \mathbf{U}_n\boldsymbol{\Sigma}_n^{1/2}$, the state matrix is $\mathbf{A} = \mathbf{O}_n^{\uparrow\dagger}\mathbf{O}_n^{\downarrow}$ (the arrows removing the last and first block rows, $\dagger$ the pseudo-inverse) and the output matrix $\mathbf{C}$ is the first block row of $\mathbf{O}_n$. The eigenvalues $\mu_i$ and eigenvectors $\boldsymbol{\psi}_i$ of $\mathbf{A}$ give the continuous-time poles $\lambda_i = f_s\ln\mu_i$, the natural frequencies $f_i = |\lambda_i|/2\pi$, the damping ratios $\zeta_{\mathrm{id},i} = -\mathrm{Re}(\lambda_i)/|\lambda_i|$ and the mode shapes $\boldsymbol{\phi}_i = \mathbf{C}\boldsymbol{\psi}_i$. Of each complex-conjugate pair only the pole with $\mathrm{Im}(\lambda)>0$ is retained, so that a mode is represented by the same shape $\boldsymbol{\phi}$, and not by its conjugate, at every model order; this matters for strongly complex modes, whose conjugate shapes are nearly orthogonal. The columns of $\mathbf{Y}$ are free-decay responses rather than random responses; their covariances are again sums of decaying exponentials with the same poles, which is the premise of the Natural Excitation Technique \cite{james1993,james1995}.

\subsection{Automated interpretation, consensus and uncertainty}

The model was computed for orders $n=1,\dots,100$. A pole at order $n$ was labeled stable if a pole at order $n+1$ satisfied
\begin{equation}
\frac{|f_i^{(n)}-f_j^{(n+1)}|}{f_i^{(n)}}\le 1\,\%, \quad
\frac{|\zeta_i^{(n)}-\zeta_j^{(n+1)}|}{\zeta_i^{(n)}}\le 5\,\%, \quad
\mathrm{MAC}\big(\boldsymbol{\phi}_i^{(n)},\boldsymbol{\phi}_j^{(n+1)}\big)\ge 0.98,
\label{eq:stab}
\end{equation}
with $\mathrm{MAC}(\mathbf{a},\mathbf{b}) = |\mathbf{a}^{\mathsf H}\mathbf{b}|^2/\big((\mathbf{a}^{\mathsf H}\mathbf{a})(\mathbf{b}^{\mathsf H}\mathbf{b})\big)$ \cite{allemang2003}, restricted to $0<f<\SI{300}{\hertz}$ and $0<\zeta_\mathrm{id}<0.2$.

The short (\SI{0.64}{\second}), highly damped responses make the stabilization diagram sensitive to the number of block rows $s$ (Eq.~\ref{eq:hankel}), and picking modes from one diagram by eye is not reproducible. Instead, the stable poles from four identifications with $s = 20, 40, 60$ and 120 were pooled and grouped by agglomerative clustering (average linkage) with the distance \cite{magalhaes2009,reynders2012auto}
\begin{equation}
d_{ij} = \frac{|f_i - f_j|}{\max(f_i, f_j)} + 1 - \mathrm{MAC}(\boldsymbol{\phi}_i,\boldsymbol{\phi}_j),
\label{eq:dist}
\end{equation}
cut at $d=0.1$. A cluster was accepted as a mode if it contained at least 15 poles and was stable in at least three of the four settings. Accepted clusters with the same dominant wavenumber (defined below) were merged if their frequencies differed by less than 1.5\,\%, or by less than 3\,\% with $\mathrm{MAC}\ge0.8$ between their representative shapes. For each mode, the natural frequency and damping are the medians over its poles, the mode shape is the medoid of the cluster, and the damping is corrected with Eq.~(\ref{eq:zcorr}).

The uncertainty due to the finite number of impacts was estimated with a nonparametric bootstrap \cite{efron1993}. In each of 200 replicates, the segments of every sector were resampled with replacement, $\mathbf{Y}$ was rebuilt and identified with $s=60$, and the stable poles were clustered as above. The resulting clusters were matched one-to-one to the accepted modes by minimizing Eq.~(\ref{eq:dist}) (Hungarian assignment); a match was accepted if the frequencies differed by less than 2\,\% and $\mathrm{MAC}\ge0.5$. The detection rate of a mode is the fraction of replicates with an accepted match, and the 2.5 and 97.5 percentiles of the matched frequencies and damping ratios give 95\,\% intervals. Modes detected in at least 80\,\% of the replicates are called robust, the others tentative.

Two quantities characterize each mode shape. The dominant circumferential wavenumber $k$ is the index of the largest coefficient of the spatial discrete Fourier transform of $\boldsymbol{\phi}$ over the 36 sectors, with the coefficients of $+k$ and $-k$ added. The modal phase collinearity \cite{pappa1993}
\begin{equation}
\mathrm{MPC} = \left(\frac{2\lambda_1}{\lambda_1+\lambda_2} - 1\right)^{2},
\label{eq:mpc}
\end{equation}
where $\lambda_1\ge\lambda_2$ are the eigenvalues of the $2\times2$ covariance matrix of the real and imaginary parts of the mean-centered shape, equals 1 for a monophase (real) mode and tends to 0 for a strongly complex one.

\subsection{Verification of the identification}
\label{sec:verification}

Two checks were made. First, a ground-truth test of the whole identification chain. Synthetic impact responses were generated at the 196 measured impact angles from a modal model consisting of the robust modes of Section~\ref{sec:results}, with their frequencies, corrected damping ratios and complex shapes; the modal amplitudes were obtained by a least-squares fit of this model to the measured $\mathbf{Y}$. To each synthetic segment the measured residual of the corresponding impact (the segment minus its sector mean) was added, with the residuals randomly reassigned among the impacts of a sector and randomly signed in each of 20 realizations. The synthetic data therefore contain the real noise, the real impact-to-impact variability and the real residual contact-patch passages, at a signal-to-noise ratio of \SI{3.8}{\decibel} in the windowed sector averages. They were processed exactly as the measurements. In six of the realizations the full bootstrap (50 replicates) was also run, to estimate how many spurious modes pass the robustness criterion. Second, the sensitivity of the results to two processing choices: the window end factor ($\beta = 1$, 5 and 20\,\%) and the number of sectors (24, 36 and 48).

%% ===================================================================
\section{Results}
\label{sec:results}

\subsection{Stabilization and identified modes}

Figure~\ref{fig:oma}(a) shows the first four singular values of the cross-power spectral matrix. The cleat excitation is concentrated below about \SI{250}{\hertz}: the mean level of the first singular value between 300 and \SI{500}{\hertz} is \SI{23}{\decibel} below its maximum between 30 and \SI{250}{\hertz}. At \SI{3.9}{\hertz} resolution and with damping of several percent the peaks of the singular values overlap, so closely spaced modes are not separated reliably by peak picking. This limitation of FDD for highly damped structures with short records was already noted in \cite{dash2020thesis}.

The stabilization diagram for $s=60$ (Fig.~\ref{fig:oma}b) shows clear columns of stable poles below \SI{200}{\hertz}; the diagrams of the other three settings are shown in the SI (Fig.~S2). Above about \SI{200}{\hertz} the stable poles split into many short families whose frequencies and shapes change with model order and setting, and none of them meets the acceptance rules. This is consistent with the excitation roll-off in Fig.~\ref{fig:oma}(a) and sets the usable band of the present cleat test to about \SI{200}{\hertz}.

Table~\ref{tab:modes} lists the thirteen accepted modes. Ten are robust, with detection rates between 92 and 100\,\%; the modes at 49.5, 166.6 and \SI{170.7}{\hertz} are tentative, with detection rates of 78, 32 and 48\,\%. All thirteen are stable in all four SSI settings, and the per-setting median frequencies of the robust modes differ by at most 1.4\,\%. The bootstrap 95\,\% intervals of the natural frequencies of the robust modes are 0.7 to \SI{3.7}{\hertz} wide (SI, Fig.~S3). Because the bootstrap uses a single setting ($s=60$) while the point estimates pool four, a few intervals are not centered on the point estimate; they should be read as the spread caused by the sampling of impacts.

\begin{table}[t]
\centering
\caption{Modal parameters of the rolling tire (30~km/h, 1334~N) identified from the virtual array. $f_n$ and $\zeta_{\mathrm{id}}$: medians over all stable poles of the mode pooled across the four SSI-Cov settings; $\Delta f_s$: range of the per-setting medians; $\zeta$: damping after removal of the exponential-window contribution (Eq.~\ref{eq:zcorr}); 95\,\% intervals and detection rate from 200 bootstrap resamplings of the cleat impacts; $k$: dominant circumferential wavenumber; MPC: modal phase collinearity of the medoid shape; $n_s$: number of settings in which the mode is stable. Modes detected in fewer than 80\,\% of the resamplings are reported as tentative.}
\label{tab:modes}
\footnotesize
\setlength{\tabcolsep}{2.6pt}
\begin{tabular}{@{}cc ccc ccc ccc l@{}}
\toprule
 & & \multicolumn{3}{c}{Natural frequency (Hz)} & \multicolumn{3}{c}{Damping ratio (\%)} & & & & \\
\cmidrule(lr){3-5}\cmidrule(lr){6-8}
Mode & $k$ & $f_n$ & 95\,\% CI & $\Delta f_s$ & $\zeta_{\mathrm{id}}$ & $\zeta$ & 95\,\% CI & MPC & $n_s$ & Det.\ (\%) & Status \\
\midrule
1 & 1 & 45.7 & [44.8, 45.9] & 0.7 & 6.3 & 3.8 & [3.5, 6.3] & 0.02 & 4/4 & 98 & robust \\
2 & 1 & 49.5 & [50.0, 50.5] & 2.1 & 7.4 & 5.1 & [4.7, 5.4] & 0.63 & 4/4 & 78 & tentative \\
3 & 1 & 53.1 & [52.8, 53.2] & 0.2 & 5.6 & 3.4 & [3.0, 3.9] & 0.47 & 4/4 & 100 & robust \\
4 & 1 & 60.9 & [60.6, 61.4] & 0.4 & 5.1 & 3.2 & [3.2, 4.3] & 0.00 & 4/4 & 97 & robust \\
5 & 2 & 110.2 & [108.9, 112.0] & 0.8 & 4.0 & 3.0 & [2.5, 4.4] & 0.21 & 4/4 & 92 & robust \\
6 & 2 & 122.1 & [120.3, 123.8] & 1.5 & 5.1 & 4.2 & [3.6, 5.0] & 0.38 & 4/4 & 100 & robust \\
7 & 3 & 132.0 & [131.5, 132.8] & 0.6 & 4.4 & 3.5 & [3.5, 4.4] & 0.20 & 4/4 & 100 & robust \\
8 & 3 & 143.1 & [142.5, 143.2] & 1.4 & 3.5 & 2.7 & [2.5, 3.0] & 0.09 & 4/4 & 98 & robust \\
9 & 3 & 146.5 & [144.1, 147.4] & 1.7 & 2.9 & 2.1 & [2.0, 2.8] & 0.32 & 4/4 & 92 & robust \\
10 & 4 & 156.3 & [155.0, 156.6] & 1.7 & 4.8 & 4.1 & [4.1, 5.0] & 0.21 & 4/4 & 100 & robust \\
11 & 4 & 166.6 & [166.0, 167.2] & 3.1 & 3.6 & 2.9 & [2.7, 3.2] & 0.13 & 4/4 & 32 & tentative \\
12 & 4 & 170.7 & [169.6, 171.4] & 1.9 & 4.3 & 3.6 & [3.5, 4.1] & 0.08 & 4/4 & 48 & tentative \\
13 & 5 & 193.1 & [192.1, 193.8] & 0.7 & 3.4 & 2.8 & [2.6, 2.9] & 0.18 & 4/4 & 100 & robust \\
\bottomrule
\end{tabular}
\end{table}

The identified damping ratios $\zeta_\mathrm{id}$ range from 2.9 to 6.3\,\% for the robust modes. Removing the window contribution (Eq.~\ref{eq:zcorr}) lowers them to 2.1--4.2\,\% (Fig.~\ref{fig:oma}c); the correction is largest for the low-frequency modes, where $\sigma_w/\omega_n$ is largest. The damping intervals are wide in relative terms, up to about 3 percentage points for the mode at \SI{45.7}{\hertz}.

\subsection{Verification}

\begin{figure}[p]
\centering
\includegraphics[width=\textwidth]{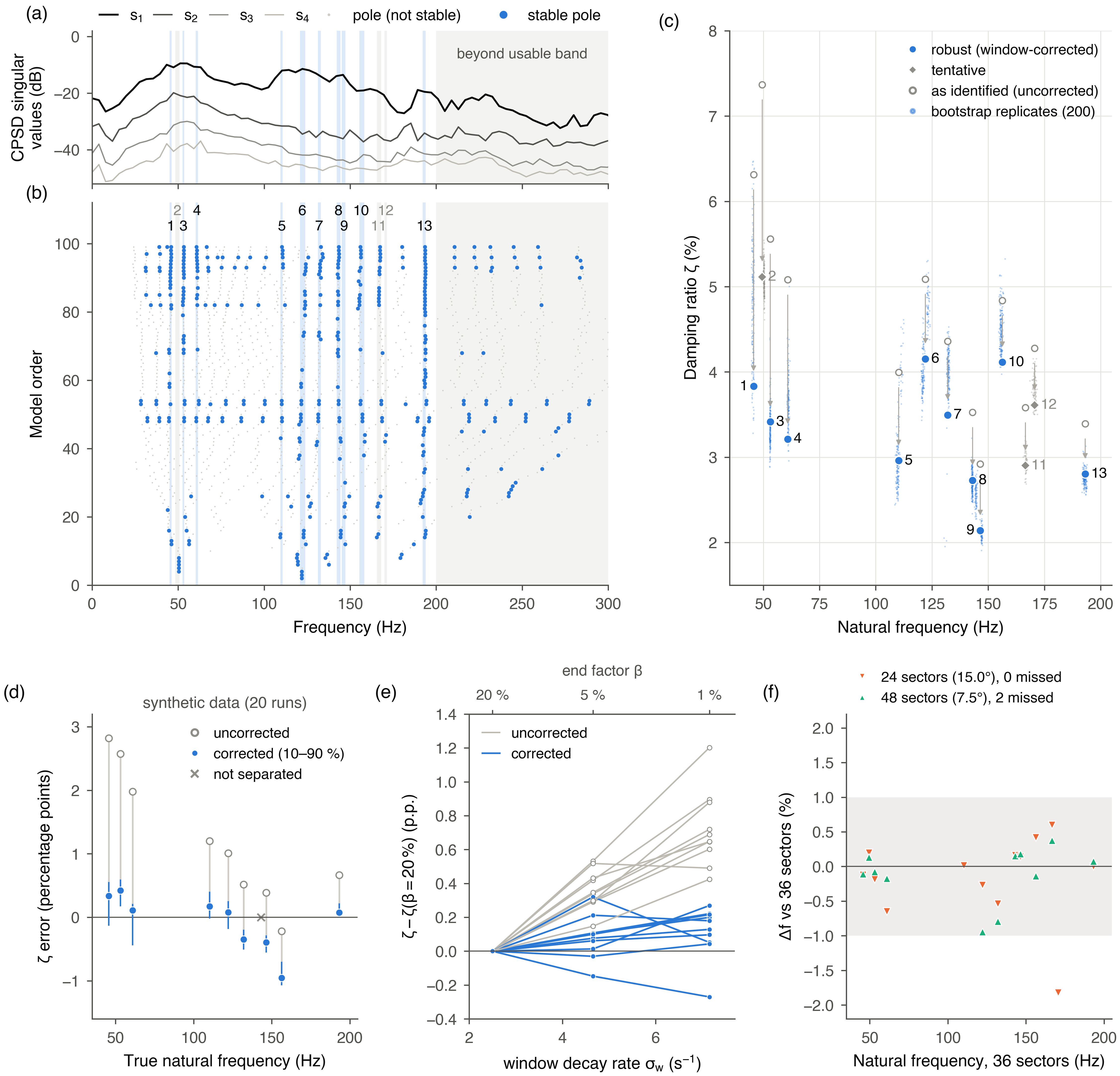}
\caption{Modal identification and its verification (dataset D1). (a) First four singular values $s_1$--$s_4$ of the cross-power spectral matrix of $\mathbf{Y}$ (resolution \SI{3.9}{\hertz}). (b) SSI-Cov stabilization diagram for $s=60$; shaded bands mark the interquartile frequency range of the accepted modes (blue: robust, grey: tentative), numbered as in Table~\ref{tab:modes}; the grey area lies above the usable band. (c) Natural frequency and damping ratio of the accepted modes: consensus estimates after the window correction of Eq.~(\ref{eq:zcorr}), estimates before correction (open circles) and 200 bootstrap replicates. (d) Synthetic test with known modes and measured noise: median damping error of the identified modes before and after the window correction (bars: 10--90\,\% over 20 realizations; cross: mode not separated from its neighbor). (e) Change of the identified and corrected damping with the window end factor $\beta$, relative to $\beta=20$\,\%, for the modes found at all three values. (f) Change of the natural frequencies when the impacts are grouped into 24 or 48 sectors instead of 36.}
\label{fig:oma}
\end{figure}

The synthetic test (Section~\ref{sec:verification}) supports the identification and the damping correction. Without noise, all ten modes were recovered, without spurious modes, and the corrected damping ratios equalled the true values to the fourth decimal. With the measured residuals added, the frequencies were recovered within 1\,\% (median errors below 0.8\,\%). Before correction the identified damping exceeded the true value by 0.4 to 2.8 percentage points, and after correction the median error was within $\pm0.4$ points for eight of the ten modes (Fig.~\ref{fig:oma}d). Two limitations appeared. The mode at \SI{156.3}{\hertz} was recovered in only 30\,\% of the realizations and its corrected damping was underestimated by about one point; and the mode at \SI{143.1}{\hertz} was never separated from its neighbor at \SI{146.5}{\hertz}, \SI{2.4}{\percent} away with the same wavenumber. The parameters of this pair should therefore be read with caution. Noise also produced spurious modes at the consensus stage (16 in six realizations), but the bootstrap criterion rejected three quarters of them: of the 48 modes classified as robust, 44 were true modes, and the four spurious ones lay within 2.4\,\% of a true mode, as fragments of it, rather than at arbitrary frequencies.

The processing choices have a limited effect (Fig.~\ref{fig:oma}e,f). Changing the window end factor from 1 to 20\,\% changes the identified damping of the ten modes found at all three values by 0.4 to 1.2 percentage points, but the corrected damping by at most 0.27 points, which confirms the window correction on the measured data. The low-frequency modes are lost at $\beta = 20$\,\%, where the weaker window leaves more of the residual contact-patch passages in $\mathbf{Y}$. With 24 sectors all thirteen modes were found again, and with 48 sectors (some of which contain a single impact) eleven; the frequencies of the matched modes changed by less than 1\,\% in most cases.

\subsection{Mode shapes and complexity}

\begin{figure}[t]
\centering
\includegraphics[width=\textwidth]{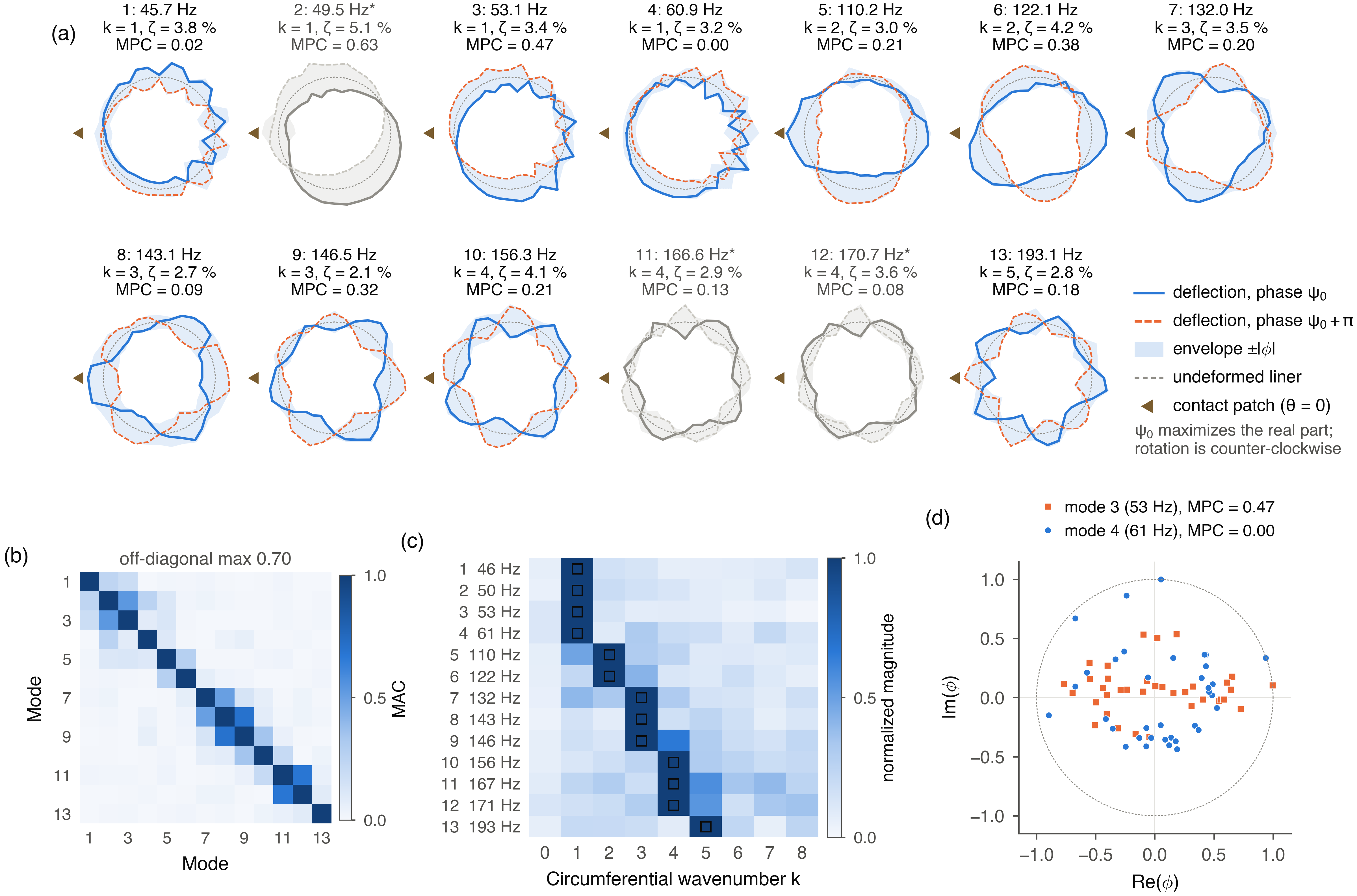}
\caption{Identified deflection patterns (dataset D1). (a) Medoid shapes of the thirteen modes on the inner liner, at the phase $\psi_0$ that maximizes the real part and at $\psi_0+\pi$, with the envelope $\pm|\boldsymbol{\phi}|$; the triangle marks the contact patch ($\theta=0$); asterisks and grey lines denote tentative modes. (b) AutoMAC matrix. (c) Normalized magnitude of the spatial Fourier transform of each shape; squares mark the dominant wavenumber $k$. (d) Complexity (Argand) plot of the most and least collinear robust modes.}
\label{fig:shapes}
\end{figure}

Figure~\ref{fig:shapes}(a) shows the medoid shape of each mode at two opposite phases, together with its envelope $|\boldsymbol{\phi}|$. The dominant wavenumber increases with frequency in a clear staircase (Fig.~\ref{fig:shapes}c): four modes with $k=1$ (45.7 to \SI{60.9}{\hertz}), two with $k=2$ (110.2 and \SI{122.1}{\hertz}), three with $k=3$ (132.0 to \SI{146.5}{\hertz}), three with $k=4$ (156.3 to \SI{170.7}{\hertz}) and one with $k=5$ (\SI{193.1}{\hertz}). The AutoMAC matrix (Fig.~\ref{fig:shapes}b) has a median off-diagonal value of 0.01; the largest values (up to 0.70) occur between modes of the same $k$, which is expected because the virtual array samples only one circumferential line at the tread center. In the bootstrap, the matched shapes of the robust modes had median MAC values between 0.81 and 0.93 with the reference shapes.

The shapes are strongly complex. The MPC of the robust modes lies between 0.00 and 0.47 (Table~\ref{tab:modes}), and in Fig.~\ref{fig:shapes}(a) the envelope $|\boldsymbol{\phi}|$ does not vanish at the nodes of the real snapshot: the nodal points move as the phase advances, that is, the deflection travels around the tire rather than standing. Figure~\ref{fig:shapes}(d) contrasts the most collinear robust mode (\SI{53.1}{\hertz}, MPC = 0.47) with the least collinear one (\SI{60.9}{\hertz}, MPC = 0.00), whose components are spread over all phases.

\subsection{Comparison with stationary modal tests}

To check the physical plausibility of the identified modes, they were compared with an EMA of the same tire at rest, inflated to \SI{241}{\kilo\pascal} (35~psi). In the unloaded case, radial frequency response functions were measured at 24 tread-center points (\SI{15}{\degree} spacing); in a loaded case (\SI{4448}{\newton}) the contact region (240--\SI{300}{\degree}) was not instrumented and 19 points were used. Modal parameters were extracted by a least-squares rational-fraction fit. For the unloaded tire, the dominant wavenumber of each mode was obtained by a least-squares fit of circumferential harmonics (SI, Fig.~S6); the fit is ill-conditioned for the loaded case with its gap in the contact region, and the mode labels of the original EMA were used instead.

\begin{figure}[t]
\centering
\includegraphics[width=\textwidth]{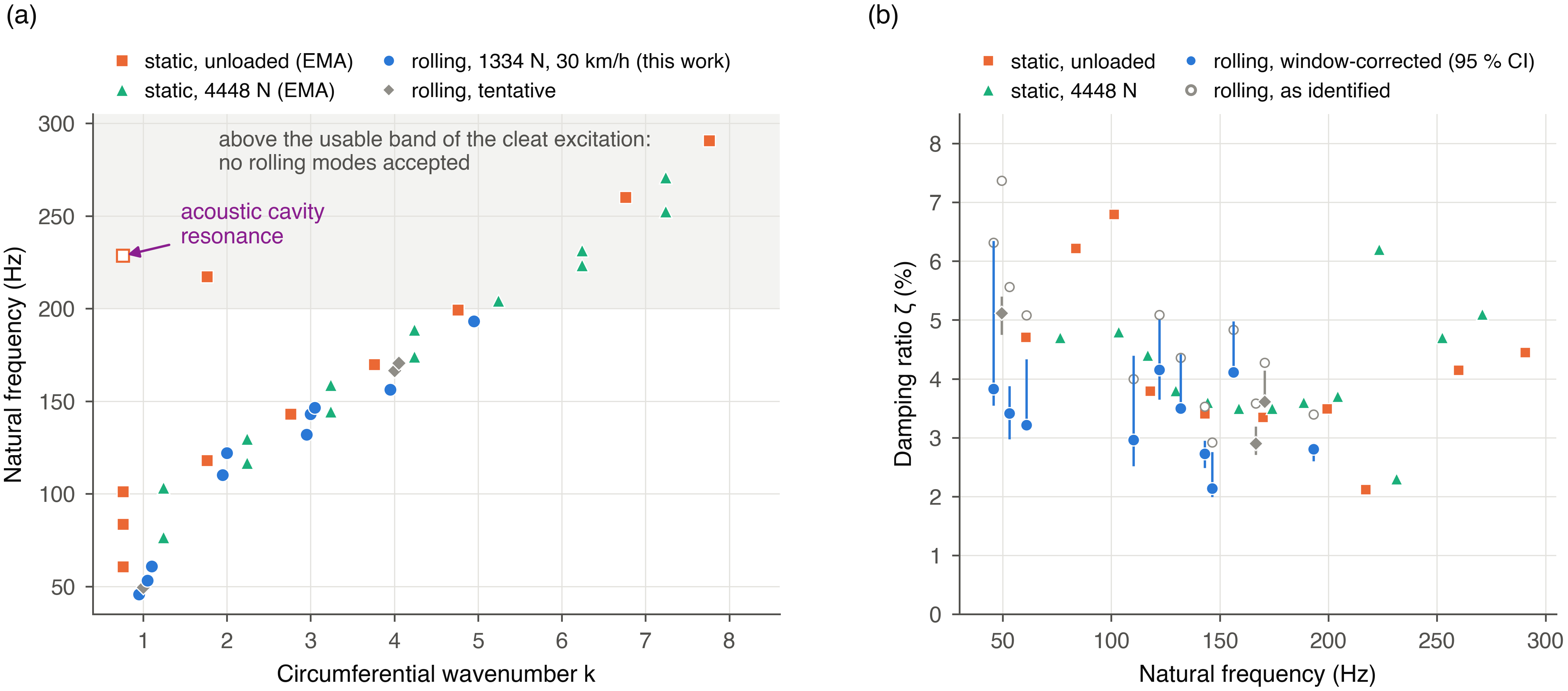}
\caption{Comparison with stationary EMA of the same tire. (a) Natural frequency versus dominant circumferential wavenumber $k$ for the rolling tire (this work; bars: bootstrap 95\,\% intervals), the unloaded static tire and the static tire under \SI{4448}{\newton}; markers are offset horizontally for legibility, and the shaded band lies above the usable bandwidth of the cleat test. (b) Damping ratio versus natural frequency; open circles show the rolling-tire values before the window correction.}
\label{fig:physics}
\end{figure}

Figure~\ref{fig:physics}(a) places the three data sets on a frequency--wavenumber diagram. For $k\ge2$ the unloaded tire shows the familiar ladder of circumferential modes (117.9, 143.0, 169.9, 199.3 and \SI{229.4}{\hertz} for $k=2$ to 6), and the rolling modes fall on the same ladder. For each $k$ from 2 to 5 the lowest rolling mode lies 3 to 8\,\% below the unloaded mode (110.2 vs.\ 117.9, 132.0 vs.\ 143.0, 156.3 vs.\ 169.9 and 193.1 vs.\ \SI{199.3}{\hertz}), while the other modes of the same $k$ lie within 4\,\% of it. Lower frequencies under rolling are consistent with the softening reported for rolling tires by Rocca et al.\ \cite{rocca2011indoor}, although the different inflation pressures (32 and 35~psi) also contribute and prevent a quantitative statement. The loaded static test, at more than three times the load of the rolling test, shows split pairs that lie mostly above the unloaded modes. The \SI{228.5}{\hertz} unloaded mode with $\zeta=0.5$\,\% and $k=1$ is the acoustic cavity resonance \cite{sakata1990}, and one unloaded mode at \SI{217}{\hertz} with dominant $k=2$ does not follow the ladder and probably belongs to a higher cross-sectional family that cannot be resolved with tread-center points alone.

The corrected damping ratios of the rolling tire (2.1--4.2\,\%) are of the same order as those of the static tests, which lie between 3.3 and 4.4\,\% for the structural modes between 110 and \SI{200}{\hertz} (Fig.~\ref{fig:physics}b). Without the window correction the rolling values would appear 0.6 to 2.5 percentage points higher, a bias that is largest exactly where the static and rolling damping are compared most often, at the first modes.

%% ===================================================================
\section{Discussion}
\label{sec:discussion}

\subsection{What the virtual array measures}

Each virtual sensor records the response of a material point of the liner, because the TCA rotates with the tire. The identification is therefore made in the rotating (Lagrangian) frame, whereas a stationary LDV measures at fixed points in space (Eulerian frame). For a rotating structure the two frames see different frequencies: gyroscopic coupling splits each wavenumber $k$ into a forward and a backward travelling wave \cite{bryan1890,huang1987}, and a wave of wavenumber $k$ observed from the fixed frame is shifted by $\pm k f_\mathrm{t}$ relative to the rotating frame. Rolling-tire frequencies obtained with the two techniques should therefore be compared only after accounting for the frame.

The travelling-wave character also explains the strong complexity of the identified shapes. Complex modes in rolling tires can arise from gyroscopic coupling, non-proportional damping, and the amplitude- and history-dependent behavior of filled rubber under cyclic deformation \cite{rocca2011}, while measurement noise and closely spaced modes can also make shapes appear complex \cite{imregun1995}. Here two of the most complex shapes belong to modes that are detected in more than 96\,\% of the bootstrap replicates (45.7 and \SI{60.9}{\hertz}), in the frequency range with the highest signal-to-noise ratio, and the bootstrap shapes are repeatable (median MAC 0.81 to 0.93), which suggests that the complexity is largely physical.

Up to four modes share the same dominant wavenumber. Within one cross-sectional family a ring-like structure has at most two modes per $k$, split by the contact patch \cite{kindt2006} and by rotation (for a thin inextensible ring the rotating-frame separation is of order $4kf_\mathrm{t}/(k^2+1)$ \cite{bryan1890}, that is, 4 to \SI{9}{\hertz} here). The additional modes therefore most likely belong to other cross-sectional families, for example modes dominated by lateral or sidewall motion, that share the circumferential wavenumber but differ in the cross-section \cite{wheeler2005}. A single radial sensor at the tread center cannot distinguish them; a second TCA off the tread center, or a model of the rolling tire, would be needed.

\subsection{Limitations}

Four limitations should be kept in mind. First, the cleat is not a white-noise source: its spectrum and the short records limit the usable band to about \SI{200}{\hertz} here, compared with the 0--\SI{500}{\hertz} band of interest for SBN. A sharper or taller cleat, a higher speed or longer runs (more impacts per sector) would extend the band. Second, the sensor leaves the contact patch while the tire is still responding: the response decays within about \SI{0.1}{\second}, during which the TCA travels about \SI{150}{\degree}. The virtual sensor at $\theta=0$ therefore does not represent a point held in the contact patch, and the shapes in Fig.~\ref{fig:shapes}(a) show no fixed node at the patch. Enforcing the contact constraint would require a model that relates the rotating and stationary frames. Third, the array samples only the radial motion along the tread center; lateral, tangential and sidewall motion, and cross-sectional mode families, are not observed. Fourth, with 4 to 8 impacts per sector the signal-to-noise ratio of the virtual array is modest: the ten robust modes explain 45\,\% of the variance of $\mathbf{Y}$, the remainder being mostly content above the usable band, the onset of the impact and the residual contact-patch passages. The synthetic test shows the consequences quantitatively: closely spaced modes of the same $k$ may not be separated, the weakest modes are not always detected, and a small fraction of the robust modes may be fragments of neighboring true modes.

\subsection{Practical considerations and relation to LDV}

Table~\ref{tab:compare} summarizes the main practical differences between a sequential LDV measurement of a rolling tire and the TCA virtual array. The two techniques address the same physical question from different frames and with different trade-offs, and they are best seen as complementary. LDV remains the reference when a dense spatial grid, the sidewall or several response directions are needed, and when the tire must not be modified. The TCA virtual array is attractive when optical access to the tread is difficult, when the tread is patterned, or when the measurement has to leave the laboratory. Its cost lies mainly in installing the sensor in the tire, which requires demounting the tire and drilling the liner, and in the limited bandwidth of the cleat excitation. Once installed, the array is recorded in a single continuous run with modest instrumentation.

\begin{table}[t]
\centering
\caption{Practical comparison of a sequential LDV measurement of a rolling tire \cite{kindt2009measurement,rocca2011indoor,rothberg2017} and the TCA virtual array of this work.}
\label{tab:compare}
\small
\begin{tabular}{@{}p{0.2\textwidth}p{0.37\textwidth}p{0.37\textwidth}@{}}
\toprule
 & Sequential LDV & TCA virtual array \\
\midrule
Quantity, frame & Surface velocity along the beam; fixed points in space & Radial acceleration of the inner liner; points moving with the tire \\
Spatial sampling & Free grid on tread and sidewall; set by the number of scan points & One circumferential line at the tread center; set by the angular binning (\SI{10}{\degree} here) \\
Acquisition & One run per point or group of points & One continuous run (\SI{128}{\second} here) \\
Access & Line of sight to the measured surface, typically via mirrors near the contact region & None; wireless link from the rotating wheel \\
Tread pattern & Speckle and dropout on rough or dark surfaces, reduced by recent laser sources and processing & Insensitive to tread pattern and surface finish \\
Tire modification & None & Insert in the liner; transmitter on the rim \\
Directions & One per beam; three with multi-beam systems & One (radial) per sensor \\
Environment & Laboratory & Laboratory; on-road use possible with an excitation time reference \\
\bottomrule
\end{tabular}
\end{table}

For on-road use the drum encoder must be replaced by another time reference for the excitation. A cleat-like event on the road, such as a joint or a dedicated bump, can be time-stamped from the TCA itself, from a wheel-hub accelerometer or from a position reference on the vehicle. The angular reference of Section~\ref{sec:reference} already depends only on the TCA, and the tire-periodic response can be removed in the same way, since order tracking needs only the contact-patch signature \cite{dash2020thesis}.

%% ===================================================================
\section{Conclusions}
\label{sec:conclusions}

A single tire cavity accelerometer on the inner liner, combined with a once-per-revolution encoder on the drum of a cleat-equipped tire-on-drum rig, was turned into a 36-element circumferential virtual array and used for operational modal analysis of a rolling tire. The method exploits the non-integer ratio between the drum and tire diameters, which places the sensor at a new angle at every cleat impact, and the contact-patch signature of the sensor itself, which provides the angular reference with \SI{1}{\degree} precision without a tire encoder.

For a slick 195/65R15 tire at \SI{30}{\kilo\meter\per\hour} and \SI{1334}{\newton}, a consensus over four SSI-Cov settings and a bootstrap over 196 impacts identified ten robust modes between 45.7 and \SI{193.1}{\hertz}, with circumferential wavenumbers 1 to 5, and three tentative ones. Correcting for the exponential window reduced the identified damping ratios by 0.6 to 2.5 percentage points, to 2.1--4.2\,\%, comparable with those of a stationary EMA of the same tire; a synthetic test with known modes and the measured noise, and the independence of the corrected damping from the window end factor, confirm this correction. The shapes are strongly complex, consistent with travelling waves in the rotating frame, and the modes follow the wavenumber ladder of the stationary tire.

The approach complements laser vibrometry: it requires installing a sensor inside the tire and is limited here to about \SI{200}{\hertz} by the cleat excitation, but it needs no optical access, is independent of the tread pattern and records the whole array in one run. Future work will address tests on treaded tires and on the road, a broader excitation and more impacts per sector to extend the usable band and the signal-to-noise ratio, a second in-tire sensor to resolve cross-sectional families, parametric studies of speed, load and pressure, and a model-based treatment of the rotating-frame measurement that enforces the contact-patch constraint.

\section*{CRediT authorship contribution statement}
\textbf{Pradosh P. Dash:} Conceptualization, Methodology, Investigation, Software, Formal analysis, Visualization, Writing -- original draft. \textbf{Pablo A. Tarazaga:} Conceptualization, Supervision, Writing -- review \& editing.

\section*{Declaration of competing interest}
The authors declare that they have no known competing financial interests or personal relationships that could have appeared to influence the work reported in this paper.

\section*{Data availability}
Data will be made available on request.

\section*{Acknowledgments}
The authors thank Prof.\ Ricardo A. Burdisso (Virginia Tech) for his guidance and supervision during the experimental work, and the Center for Tire Research (CenTiRe) for providing the test facilities and financial support for this study.

\bibliographystyle{elsarticle-num}
\bibliography{references}

\clearpage
\section*{Supplementary Information}
\addcontentsline{toc}{section}{Supplementary Information}
\setcounter{figure}{0}\setcounter{table}{0}
\renewcommand{\thefigure}{S\arabic{figure}}
\renewcommand{\thetable}{S\arabic{table}}

\noindent This supplement supports the main text. Section~S1 details the signal-processing chain, Section~S2 the identification results for all settings, Section~S3 the repeatability of the cleat impacts, Section~S4 the angular reference in the second test campaign and Section~S5 the wavenumber analysis of the stationary modal test.

\subsection*{S1. Signal-processing chain}

\begin{figure}[h]
\centering
\includegraphics[width=\textwidth]{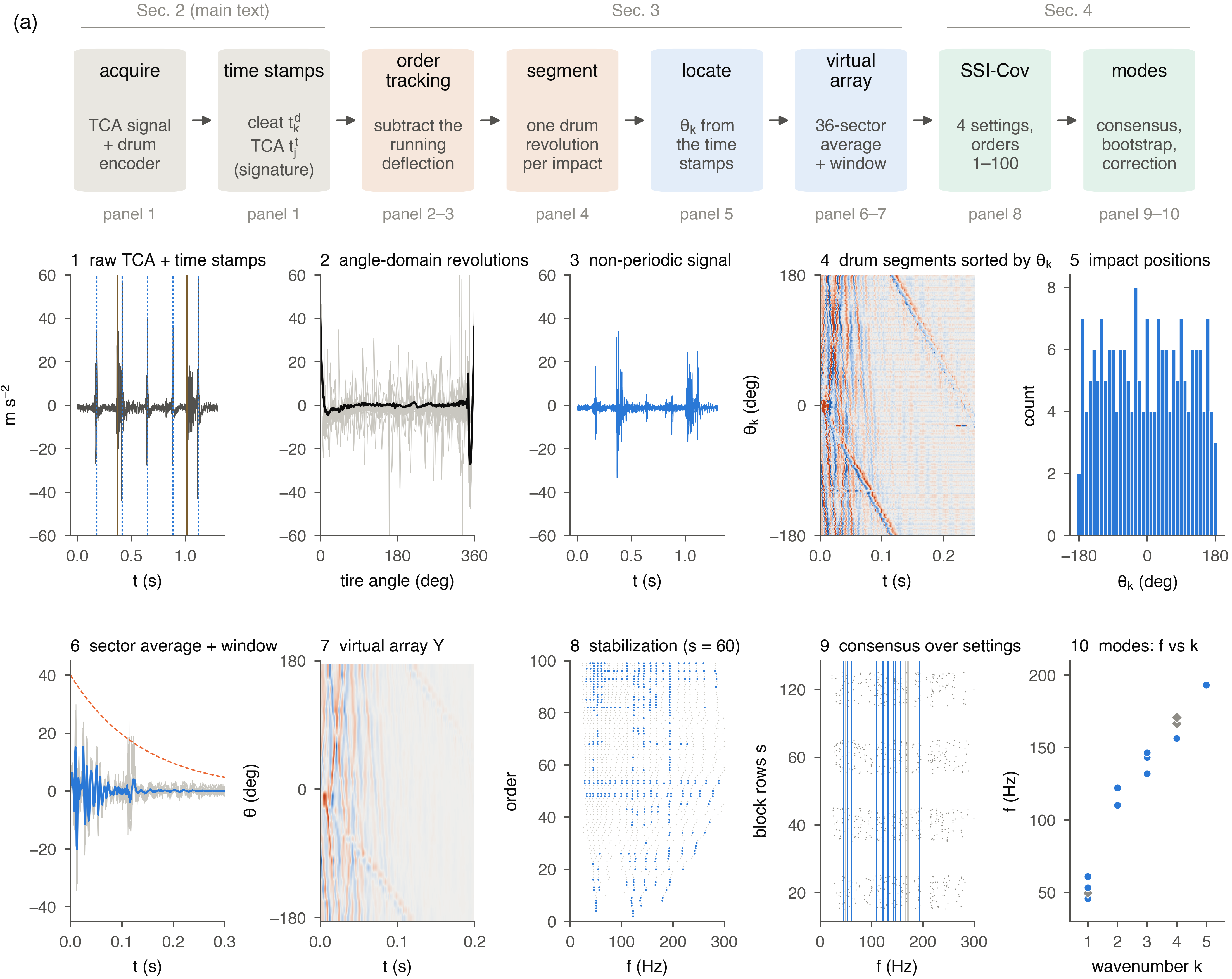}
\caption{The signal-processing chain. (a) Flowchart of the processing steps, with the section of the main text in which each step is described and the panel below in which it is illustrated. Panels 1--10: the processing applied to dataset D1; panel numbers follow the list above.}
\label{fig:S1}
\end{figure}

Figure~\ref{fig:S1}(a) summarizes the processing chain described in Sections~2--4 of the main text, and the panels below it show the data of dataset D1 at every step.
\begin{enumerate}\itemsep0pt
\item The raw TCA signal is time-stamped with the cleat passages (drum encoder) and the TCA passages through the contact patch (trailing-edge peak of the TCA signature).
\item Each tire revolution is resampled to 450 points; the average over all revolutions is the tire-periodic (running-deflection) response.
\item Its subtraction leaves the non-periodic signal, dominated by the cleat transients.
\item The non-periodic signal is cut into one drum revolution per impact (1300 points, $f_s = \SI{2018.5}{\hertz}$); sorted by the TCA position $\theta_k$ the segments already show the structure of the virtual array, including the residual contact-patch passage as an oblique trace.
\item The 196 impact positions cover all 36 sectors.
\item The segments of each sector are averaged and multiplied by an exponential window (end factor 1\,\%).
\item The 36 windowed averages form the virtual-array matrix $\mathbf{Y}$.
\item SSI-Cov is computed for model orders 1--100 and four numbers of block rows; the stabilization diagram for $s=60$ is shown.
\item Stable poles of the four settings are pooled and clustered; vertical lines mark the accepted modes (blue: robust, grey: tentative).
\item Natural frequency versus dominant circumferential wavenumber of the accepted modes.
\end{enumerate}

\subsection*{S2. Identification results for all settings}

\begin{figure}[h]
\centering
\includegraphics[width=\textwidth]{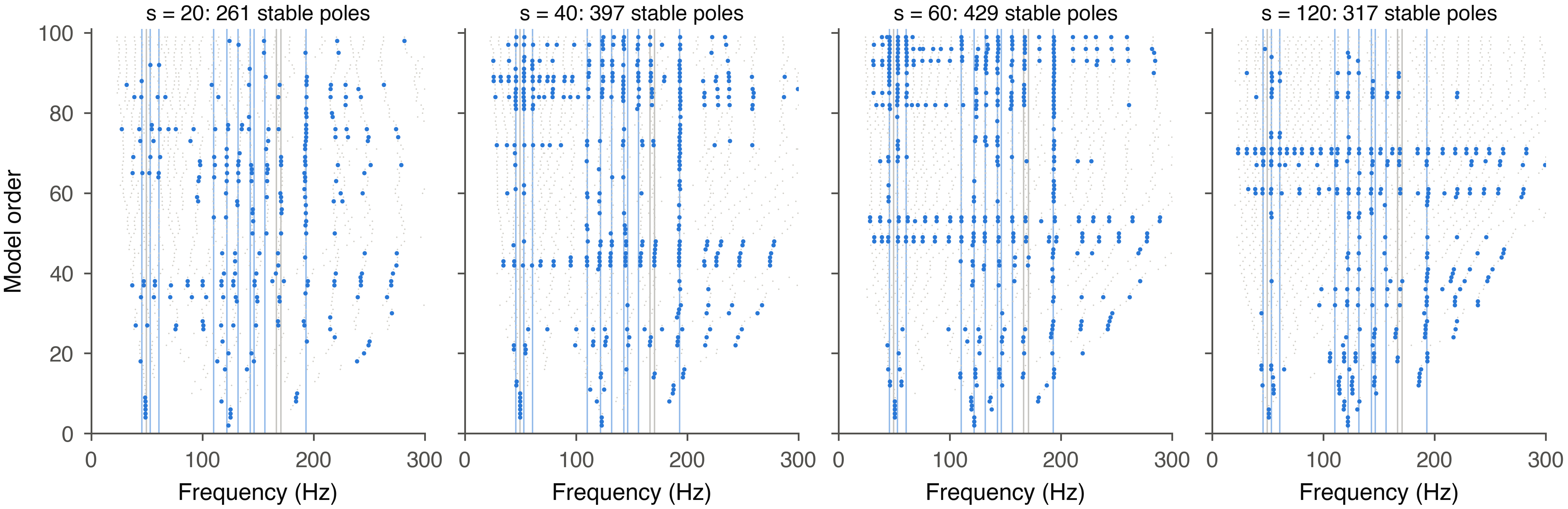}
\caption{Stabilization diagrams for $s = 20$, 40, 60 and 120 block rows (grey: poles; blue: stable poles; vertical lines: accepted modes, blue robust and grey tentative).}
\label{fig:S3}
\end{figure}

\begin{figure}[p]
\centering
\includegraphics[width=\textwidth]{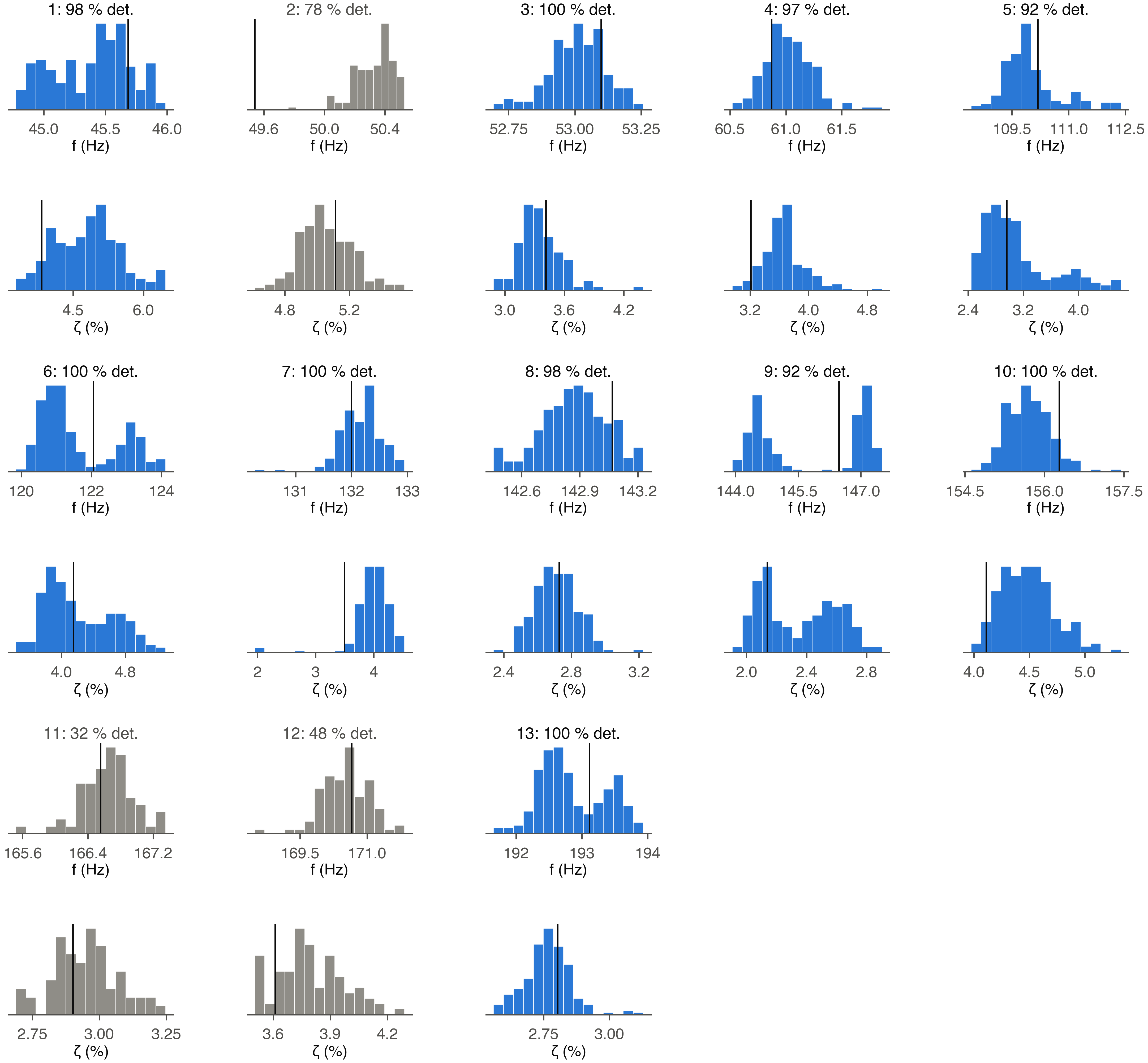}
\caption{Bootstrap distributions (200 resamplings of the impacts within sectors) of the natural frequency and the window-corrected damping ratio of each accepted mode, with its detection rate; vertical lines: consensus estimates; grey: tentative modes.}
\label{fig:S4}
\end{figure}

Figure~\ref{fig:S3} shows the stabilization diagrams of the four SSI-Cov settings; vertical lines mark the accepted modes. Table~\ref{tab:S1} lists the per-setting medians of the natural frequency and damping ratio of each accepted mode. Figure~\ref{fig:S4} shows the bootstrap distributions of the natural frequencies and window-corrected damping ratios; the vertical line is the consensus estimate. Bimodal distributions (modes 6 and 9) indicate that the bootstrap replicates sometimes identify a neighboring mode of the same wavenumber; together with the synthetic test this is why the parameters of the pair at 143.1 and \SI{146.5}{\hertz} should be read with caution. The bootstrap distribution of mode 2 (tentative) does not contain the consensus estimate, which reflects the difference between the pooled settings and the single setting used in the bootstrap.

\begin{table}[h]
\centering
\caption{Per-setting estimates of the accepted modes: median natural frequency (Hz) and window-corrected damping ratio (\%) of the stable poles obtained with $s = 20, 40, 60$ and 120 block rows, number of stable poles per setting, and median bootstrap MAC of the matched shapes with the reference shape.}
\footnotesize
\setlength{\tabcolsep}{3pt}
\begin{tabular}{@{}cc cccc cccc cc@{}}
\toprule
 & & \multicolumn{4}{c}{$f_n$ (Hz) for $s$ =} & \multicolumn{4}{c}{$\zeta$ (\%) for $s$ =} & & \\
\cmidrule(lr){3-6}\cmidrule(lr){7-10}
Mode & $f_n$ & 20 & 40 & 60 & 120 & 20 & 40 & 60 & 120 & poles & MAC$_\mathrm{boot}$ \\
\midrule
1 & 45.7 & 45.2 & 45.7 & 45.6 & 45.8 & 3.2 & 5.3 & 3.4 & 4.2 & 2/11/16/4 & 0.92 \\
2 & 49.5 & 48.7 & 49.5 & 50.4 & 50.7 & 5.4 & 4.8 & 5.1 & 5.7 & 6/7/5/3 & 0.98 \\
3 & 53.1 & 53.0 & 53.0 & 53.1 & 53.2 & 3.7 & 3.3 & 3.4 & 3.8 & 1/15/22/13 & 0.93 \\
4 & 60.9 & 61.0 & 60.9 & 61.0 & 60.6 & 3.0 & 3.5 & 3.2 & 2.6 & 1/10/13/7 & 0.90 \\
5 & 110.2 & 110.4 & 109.7 & 110.2 & 110.4 & 2.3 & 3.0 & 2.8 & 2.9 & 4/7/5/2 & 0.91 \\
6 & 122.1 & 121.7 & 123.1 & 122.0 & 122.3 & 3.2 & 4.3 & 4.5 & 4.1 & 10/21/25/14 & 0.91 \\
7 & 132.0 & 132.2 & 132.2 & 132.4 & 131.7 & 1.7 & 3.8 & 4.2 & 3.3 & 8/16/21/12 & 0.89 \\
8 & 143.1 & 144.2 & 142.8 & 143.0 & 143.8 & 2.9 & 2.7 & 2.7 & 2.8 & 12/26/28/15 & 0.90 \\
9 & 146.5 & 146.9 & 146.5 & 146.9 & 145.3 & 2.2 & 2.3 & 2.1 & 2.1 & 3/4/4/7 & 0.84 \\
10 & 156.3 & 157.6 & 156.0 & 155.9 & 157.3 & 2.6 & 4.4 & 4.2 & 4.0 & 9/18/17/7 & 0.86 \\
11 & 166.6 & 168.5 & 166.6 & 166.4 & 165.4 & 3.3 & 2.9 & 2.9 & 2.7 & 4/3/4/5 & 0.83 \\
12 & 170.7 & 170.7 & 170.7 & 170.5 & 168.8 & 3.8 & 3.6 & 4.0 & 3.4 & 6/7/2/3 & 0.83 \\
13 & 193.1 & 192.7 & 193.0 & 193.4 & 192.9 & 3.1 & 3.0 & 2.7 & 2.7 & 35/51/50/26 & 0.81 \\
\bottomrule
\end{tabular}
\label{tab:S1}
\end{table}

\subsection*{S3. Repeatability of the cleat impacts}

\begin{figure}[h]
\centering
\includegraphics[width=\textwidth]{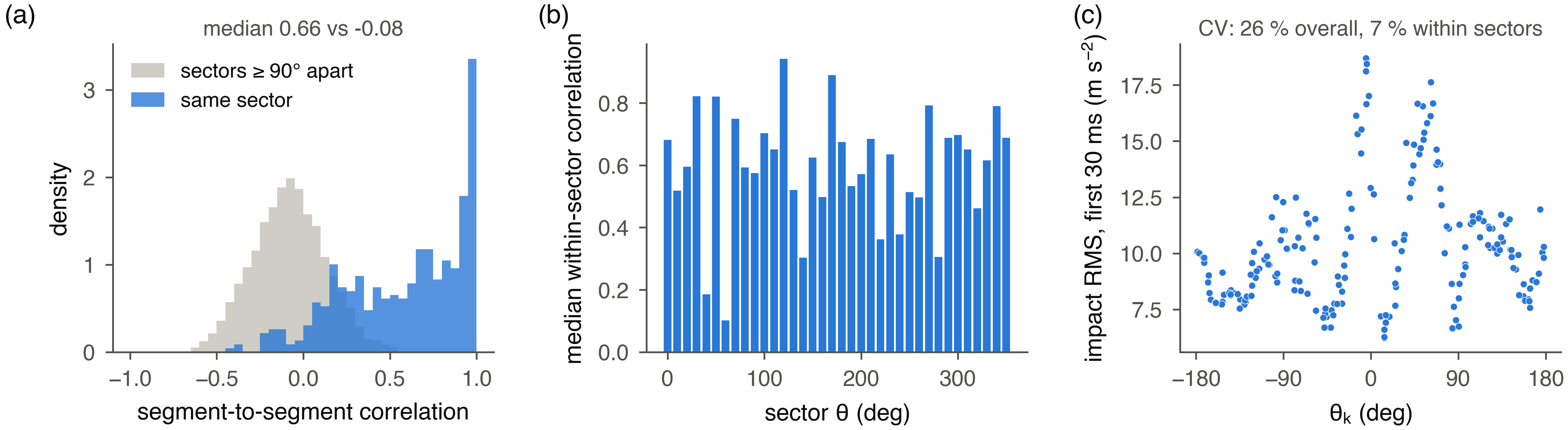}
\caption{Repeatability of the cleat-impact response (dataset D1). (a) Distribution of segment-to-segment correlations within sectors and between sectors at least \SI{90}{\degree} apart. (b) Median within-sector correlation per sector. (c) RMS of the impact response in the first \SI{30}{\milli\second} versus the TCA position.}
\label{fig:S5}
\end{figure}

Averaging within a sector assumes that the cleat input and the tire state are the same at every impact. Figure~\ref{fig:S5}(a) compares the correlation between segments of the same sector with that between segments of sectors at least \SI{90}{\degree} apart (band 30--\SI{220}{\hertz}, first \SI{0.1}{\second}). Segments of the same sector are clearly similar (median 0.66) while distant sectors are uncorrelated (median $-0.08$). The within-sector correlation is similar for all sectors (Fig.~\ref{fig:S5}b). The energy of the impact response in the first \SI{30}{\milli\second} varies strongly with the position of the sensor (Fig.~\ref{fig:S5}c), with maxima close to the contact patch, but only by 7\,\% (median coefficient of variation) within a sector, which indicates a repeatable cleat input.

\subsection*{S4. Angular reference in the second campaign}

\begin{figure}[h]
\centering
\includegraphics[width=0.6\textwidth]{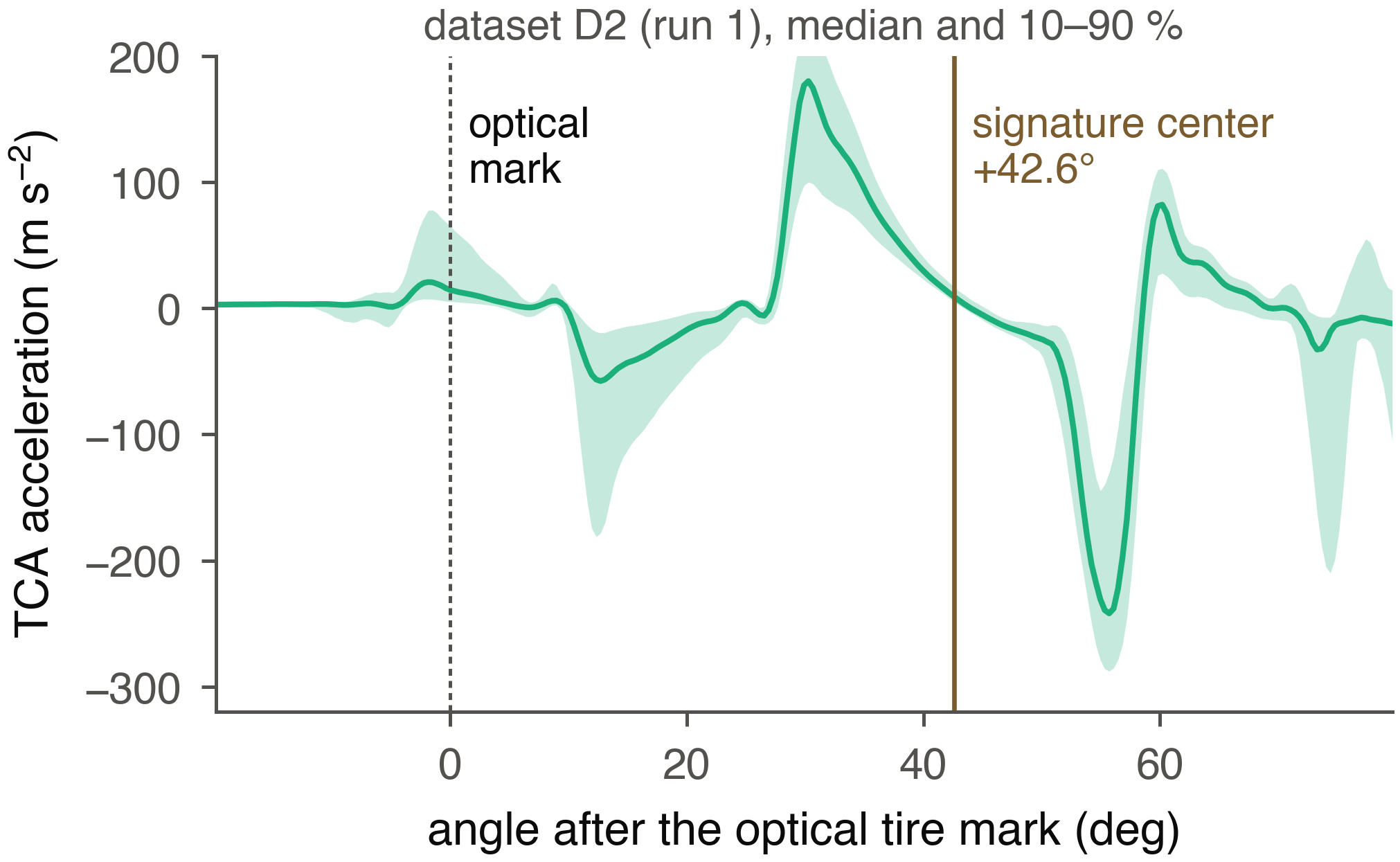}
\caption{TCA signal of dataset D2 (run 1) synchronously averaged on the optical tire encoder (median and 10--90\,\% band); the vertical line marks the center of the contact-patch signature.}
\label{fig:S6}
\end{figure}

In dataset D2 an optical encoder on the tire sidewall provided an independent once-per-revolution signal. Figure~\ref{fig:S6} shows the TCA signal synchronously averaged on this optical mark. The contact-patch signature (here a positive peak at the leading edge and a negative peak at the trailing edge; the polarity is reversed with respect to D1) is centered \SI{42.6}{\degree} after the optical mark, an offset most likely due to the position of the reflective tape relative to the TCA. The midpoint of the signature fluctuates by \SI{1.0}{\degree} (standard deviation) relative to the optical mark (main text, Fig.~2e). In D2 the TCA signature spans about \SI{25}{\degree} between its peaks, against \SI{11}{\degree} in D1; the reason for the difference is not known (the two campaigns used different data acquisition settings), and it does not affect the use of the midpoint as reference. The smaller cleat of D2 excited the tire too weakly, relative to the residual running-deflection response, for the modal identification.

\subsection*{S5. Wavenumber analysis of the stationary modal test}

\begin{figure}[h]
\centering
\includegraphics[width=0.55\textwidth]{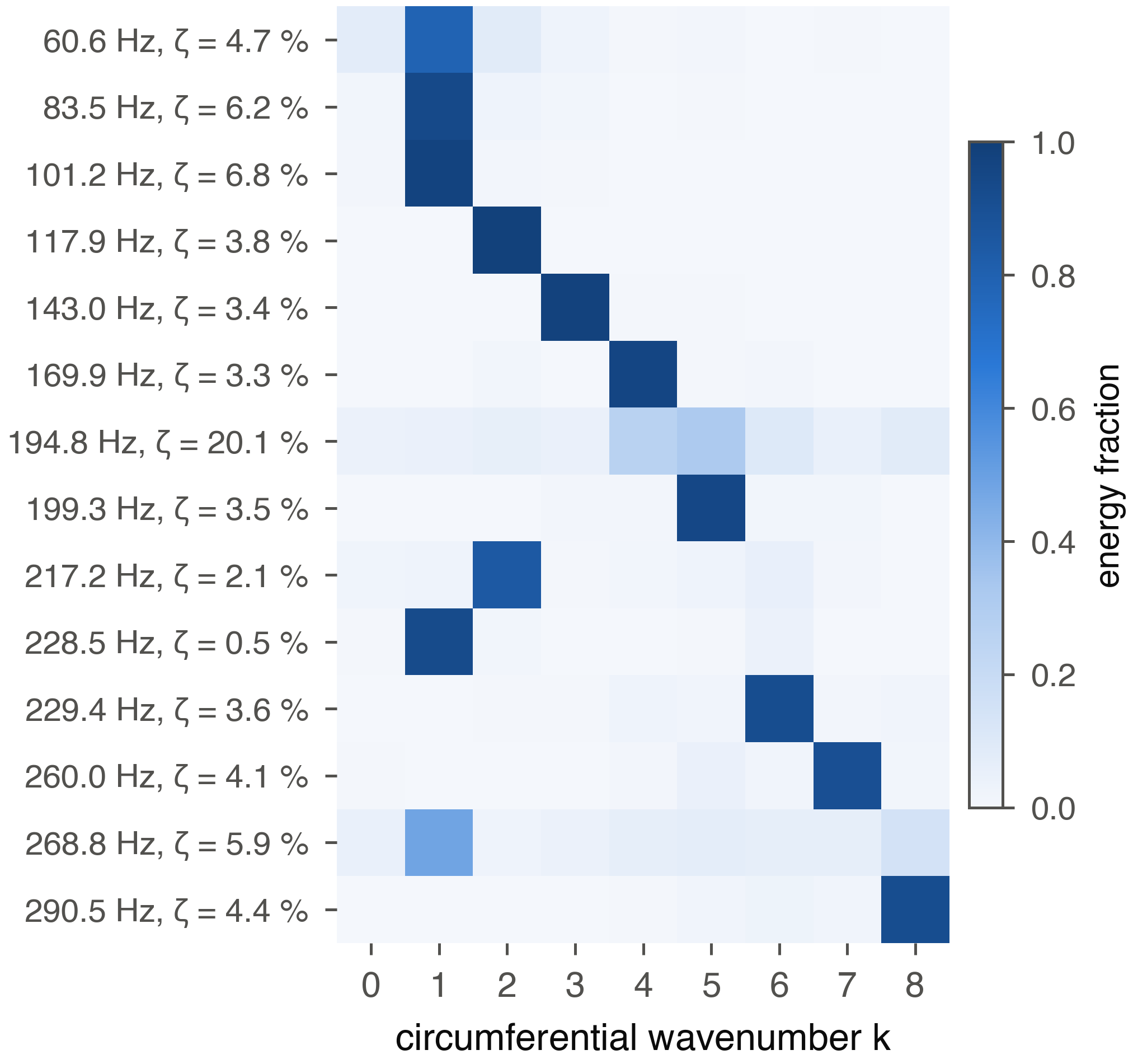}
\caption{Fraction of modal energy per circumferential wavenumber for the modes of the unloaded static EMA below \SI{300}{\hertz}.}
\label{fig:S7}
\end{figure}

For the unloaded static EMA (24 tread-center points, \SI{15}{\degree} spacing) the circumferential content of each mode shape was obtained by a least-squares fit of the harmonics $\mathrm{e}^{\mathrm{i}k\theta}$, $k=-8,\dots,8$, and the energy of $+k$ and $-k$ was added (Fig.~\ref{fig:S7}). Most structural modes have more than 90\,\% of their energy in one wavenumber. The mode at \SI{228.5}{\hertz}, with 0.5\,\% damping and $k=1$, is the acoustic cavity resonance; the modes at 194.8 and \SI{268.8}{\hertz} have no dominant wavenumber and were not used.

\end{document}